\documentclass[12pt]{ronbun}

\usepackage{amsmath,amssymb}
\usepackage{cite}
\usepackage{bm}
\usepackage{dcolumn}

\newcommand{\nn}{\nonumber}
\newcommand{\e}{{\rm e}}

\newcommand{\del}{\delta}

\newcommand{\al}{\alpha}

\renewcommand{\th}{\theta}

\newcommand{\1}{\mathbb I}
\newcommand{\CC}{\mathcal C}
\newcommand{\CM}{\mathcal M}
\newcommand{\CO}{\mathcal O}
\newcommand{\CP}{\mathcal P}

\numberwithin{equation}{section}

\begin{document}

\begin{flushright}
\parbox{4.2cm}
{KEK-TH-922 \hfill \\
{\tt hep-th/0310228}
 }
\end{flushright}

\vspace*{1.1cm}

\begin{center}
 \Large\bf D-branes of Covariant AdS Superstrings 
\end{center}
\vspace*{1.5cm}
\centerline{\large Makoto Sakaguchi$^{a}$ and Kentaroh Yoshida$^{b}$}

\begin{center}
\emph{Theory Division, High Energy Accelerator Research 
Organization (KEK),\\
Tsukuba, Ibaraki 305-0801, Japan.} 
\\
\vspace*{1cm}
$^{a}$Makoto.Sakaguchi@kek.jp
~~~~
 $^{b}$kyoshida@post.kek.jp
\end{center}

\vspace*{2cm}

\centerline{\bf Abstract}
  
\vspace*{0.5cm}

We consider D-branes of open superstrings in the AdS$_5\times S^5$
background. The possible configurations of D-branes preserving half of
supersymmetries are classified by analyzing the $\kappa$-invariance
of an open superstring
in a covariant manner.  We also revisit the
classification of D-branes in the pp-wave background. It is shown that
Penrose limits of the possible D-branes in the AdS$_5\times S^5$ give
all of the D-branes in the pp-wave. In addition, a 1/4 supersymmetric
D-string, which is related to the D-string preserving 8 dynamical
supersymmetries in the pp-wave, is presented.  We also discuss 
the relation between our result 
and the AdS branes in a brane probe analysis.

\vfill

\thispagestyle{empty}
\setcounter{page}{0}

\newpage 

\section{Introduction}

The discovery of D-branes \cite{Pol} has played an important role for
revealing non-perturbative aspects of superstring theories and M-theory
\cite{BFSS}. Recent interest in studies of D-branes is the
classification of possible configurations in non-trivial backgrounds.
Among other non-trivial backgrounds, the maximally supersymmetric type
IIB pp-wave background \cite{BFHP1} has attracted great interests
because the Green-Schwarz superstring on this background was shown to be
exactly solvable \cite{M,MT}.  The configurations of D-branes on pp-wave
backgrounds were examined in many works
\cite{DP,BP,SMT,Bain,BGG,Baryon,Michi,SY4,BPZ,Taka,HS3,Yamag,M2,
HPS,KLY,SSY,Kumar}.  In particular, covariant analyses of
D-branes of an open string in pp-waves were given in \cite{BPZ,HPS} by
applying a method developed by Lambert and West \cite{LW}.

It is also possible to apply this method for analyzing Dirichlet branes
of an open supermembrane.  In the eleven-dimensional
flat spacetime, Dirichlet $p$-branes
were shown to be allowed for the values $p=1,5,9$ \cite{EMM,dWPP}.  The
$p=5$ case corresponds to M5-brane, while the $p=9$ case describes the
end-of-world 9-brane in the Horava-Witten theory \cite{HW}. Studies 
of open supermembrane theories are expected to promote our understanding
of features of M-branes. 

Dirichlet branes of an open supermembrane on the maximally
supersymmetric pp-wave background
(Kowalski-Glikman solution)\footnote{Supermembrane on the pp-wave
background is related to the pp-wave matrix model \cite{BMN} through the
matrix regularization \cite{dWHN}. This relation in the pp-wave case is
discussed in \cite{DSR,SY1,SY2}.}  were classified in
\cite{SaYo:pp,SY1}.  In the work \cite{SaYo:pp}, we provided a covariant
classification of the possible D-brane configurations by combining
methods in Refs.\,\cite{EMM,dWPP} and Ref.\,\cite{LW}. We found 1/2
supersymmetric M5-brane and 9-brane configurations sitting at the origin
of the pp-wave.  These configurations are not 1/2 supersymmetric outside
the origin while 1-brane configurations are 1/2 supersymmetric at and
outside the origin. Also, 1/2 supersymmetric 9-brane configurations
obtained in Ref.\,\cite{SaYo:pp} are expected to be available for a
study of a heterotic matrix model \cite{Motl}.  If we consider a double
dimensional reduction of open supermembrane on the pp-wave background to
a type IIA string theory \cite{SY4} through a compactification of
one of transverse dimensions, then M5- and 9-branes in eleven dimensions
wrapping an $S^1$-circle become possible configurations of D4-branes and
D8-branes in the resulting type IIA pp-wave background
\cite{SY4,HS3,HPS,SSY}. This correspondence strongly supports 
the weak-strong 
duality between M-theory and type IIA string theory even in pp-wave backgrounds,
and thus we can say that a consistency of superstring theories on a
pp-wave has been checked.

In our previous work \cite{SaYo:ads}, Dirichlet branes of an open
supermembrane in the AdS$_{4/7}\times S^{7/4}$ backgrounds were
examined, and possible configurations were classified.  It was shown
that under the Penrose limit \cite{P} possible Dirichlet branes are
mapped one-to-one to those on the pp-wave background, as expected from
the fact that the Kowalski-Glikman solution \cite{KG} 
can be obtained from the AdS$_{4/7}\times S^{7/4}$ backgrounds \cite{BFHP2}.
The allowed Dirichlet branes are also related to AdS 
branes in the AdS$_{4/7}\times S^{7/4}$ backgrounds.

Motivated by our previous works \cite{SaYo:pp,SaYo:ads}, we will
consider D-branes of an AdS string by analyzing the covariant Wess-Zumino
term. First, we classify D-brane configurations of an open superstring
in the AdS$_5\times S^5$ background.
The possible D-brane
configurations can be compared to the AdS brane configurations. Secondly,
we extend the covariant classification of D-branes in the pp-wave
\cite{BPZ} by including the cases that the null directions ($+,-$) are
Dirichlet ones.  Thirdly, we examine Penrose limits of the possible
D-branes in the AdS$_5\times S^5$ background, and directly show that all of
D-branes in the pp-wave are realized in these limits.
In addition, we present an example of 1/4 supersymmetric D-string
configurations in the AdS$_5\times S^5$ background, which preserves 1/4
supersymmetries even outside the origin. 
This configuration reduces
to the 1/4 supersymmetric D-string in the pp-wave background
found in the work \cite{BPZ} through the Penrose limit.
From the viewpoint of embedded
branes in the AdS$_5\times S^5$ background, this D-string may
possibly correspond to
$R\times S^1$\,.

The organization of our paper is as follows: In Section 2, we introduce
the action of an open superstring on the AdS$_5\times S^5$ background,
and the
covariant Wess-Zumino term is presented.  In section 3, we classify
possible 1/2 supersymmetric D-brane configurations of the AdS string by
investigating the vanishing conditions of the $\kappa$-variation surface
terms of the covariant Wess-Zumino term. In addition, we find a 1/4
supersymmetric
D-string configuration. This D-string preserves a quarter of
supersymmetries both at and outside the origin.  In section 4, we
classify D-brane configurations preserving half of supersymmetries in
the pp-wave background which include the cases that the null directions
($+,-$) satisfy the Dirichlet conditions.  In section 5 we consider the
Penrose limit of the result obtained in section 3. All of the resulting
configurations after the Penrose limit are shown to be included in the
list presented in section 4. We also discuss the relation between our
result and AdS branes in the AdS$_5\times S^5$ background. Section 5 is devoted
to a conclusion and discussions. In appendix A, our notation and
convention are summarized.

\section{Covariant Wess-Zumino Term for the AdS String}

In this section, we will introduce the Green-Schwarz action of a
superstring on the AdS$_5\times S^5$ background (called AdS string
below), and explain some characteristics relevant to our later
considerations.  In this paper, we do not utilize any specific gauges
but discuss in a covariant fashion, because the covariant expression of
the Wess-Zumino term is available for the covariant classification of
D-branes of the AdS string.

First of all, the action of AdS string we consider is written as 
\begin{eqnarray}
\label{S}
S &=& \int\!\!d^2\sigma\,\Bigl[\mathcal{L}_{\rm NG} + 
\mathcal{L}_{\rm WZ}\Bigr]\,, \quad \mathcal{L}_{\rm NG} = -\sqrt{-g(X,\th)}
\,. 
\end{eqnarray}
The Nambu-Goto part of this Lagrangian is represented in terms of the 
induced metric $g_{ij}$, which is given by (For notation and convention,
see Appendix A)
\begin{eqnarray}
g_{ij} = E_i^ME^N_j G_{MN} = E_i^AE_j^B\eta_{AB}\,, \quad 
g = \det g_{ij}, \quad 
E^A_i=\partial_iZ^{\hat M}E_{\hat M}^A\,,
\end{eqnarray}
where $Z^{\hat M}=(X^M,\theta^{\bar \alpha})$ and
$E_{\hat M}^A$ are supervielbeins of the AdS$_5 \times S^5$ background.  
For D-strings, $g$ is replaced with $\det(g_{ij}+{\mathcal F}_{ij})$
where ${\mathcal F}$ is defined by
${\mathcal F}=dA-B$ with the Born-Infeld $U(1)$ 
gauge field $A$
and the pull-back of the NS-NS two-form 
$B$.

The Wess-Zumino term is described as\footnote{
Alternative superstring actions 
have been proposed in \cite{alternative:AdS}
for superstrings in the AdS background and 
in \cite{alternative;PP, Berkovits} for those in the pp-wave background.
}
\begin{eqnarray}
\label{WZ}
\mathcal{L}_{\rm WZ} &=& -2i\int_0^1
\!\!dt\,\widehat{E}^{A}\bar{\th}\Gamma_A\sigma 
\widehat{E}\,,
\end{eqnarray}
where $\widehat{E}^A \equiv E^A(t\th)$ and $\widehat{E}^{\al}\equiv 
E^{\al}(t\th)$.
When we consider a fundamental string (F-string), the matrix $\sigma$ 
is given by $\sigma_3$\,. If we consider a D-string, then $\sigma$ is
represented by $\sigma_1$\,. Since we would like to discuss boundary
surfaces for both of fundamental string and D-string, we do not
explicitly fix $\sigma$ in our consideration.    
It is known that $\kappa$-invariance of the action
is equivalent to supergravity equations of motion \cite{kappa1,kappa2}.
In the case of an open string,
the $\kappa$-variation leads to surface terms.
One can show that 
the surface terms originating from 
$S_{\textrm{NG}}$
vanish, 
because they include 
$\delta_\kappa Z^{\hat M}E_{\hat M}^A=\delta_\kappa E^A=0$\,,
by using the definition of the 
$\kappa$-variation $\delta_\kappa E^A=0$.
In addition, the $\kappa$-variation of $\mathcal F$
does not lead to surface terms because
$\delta_\kappa \mathcal F$ does not
include a derivative of the $\kappa$-variation of a field.
Thus non-vanishing surface terms originate from $S_{\rm{WZ}}$ only.

In this section, we consider an open string in the AdS$_5\times S^5$
background. The concrete expressions of the supervielbein and the spin
connection are given in Appendix B where the coset construction is
briefly reviewed for this background.  We examine the $\kappa$-variation
surface terms up to and including the fourth order of $\theta$ in this
paper\footnote{It is expected that the fourth order analysis is
 sufficient as we will briefly comment later.}.  
After short calculation, the Wess-Zumino term can be 
rewritten
as
\begin{eqnarray}
S_{\rm WZ} &=& S_{\rm WZ}^0 + S_{\rm WZ}^{\rm spin} + S_{\rm
 WZ}^{\mathcal{M}}\,, \nn 
\end{eqnarray}
where each part in the above decomposition is defined as, respectively,  
\begin{eqnarray}
\label{WZ:0}
S_{\rm WZ}^0 &\equiv& -2i\int\!\! d^2\sigma\,\epsilon^{ij}
\Biggl[
-\frac{1}{2}\bar{\th}\Gamma_A\sigma \left(\partial_i\th +
\frac{\lambda}{2}\widehat{\Gamma}_B i\sigma_2\th\partial_i X^{M}e_M^B\right)
\partial_j X^N e_N^A  \\
&& + \frac{i}{4}\bar{\th}\Gamma^A\left(\partial_i\th + 
\frac{\lambda}{2}\widehat{\Gamma}_B i\sigma_2\th\partial_i X^M e_M^B \right)
\cdot \bar{\th}\Gamma_A\sigma\left(\partial_j\th + 
\frac{\lambda}{2}\widehat{\Gamma}_C	i\sigma_2\th \partial_jX^Ne_N^C  \right) 
\Biggr]\,, \nn \\
\label{WZ:spin}
S_{\rm WZ}^{\rm spin} &\equiv& -2i\int\!\!d^2\sigma\,
\epsilon^{ij}
\Biggl[
-\frac{1}{8}\bar{\th}\Gamma_A\sigma\Gamma_{BC}\th\omega_M^{BC}
\partial_i X^M\partial_j X^N e_N^A \\
&& + \frac{i}{16}\bar{\th}\Gamma^A\Gamma_{BC}\th\omega_M^{BC}\partial_iX^M
\cdot\bar{\th}\Gamma_A\sigma\left(
\partial_j\th + \frac{\lambda}{2}\widehat\Gamma_D
i\sigma_2\th\partial_j X^N e_N^D
\right) \nn \\
&& + \frac{i}{16} \bar{\th}\Gamma^A\left(
\partial_i\th + \frac{\lambda}{2}\widehat{\Gamma}_B i\sigma_2 \th \partial_i
X^M e_M^B\right)\cdot\bar{\th}\Gamma_A\sigma\Gamma_{CD}\th
\omega_N^{CD}\partial_j X^N \nn \\ 
&& + \frac{i}{64}\bar{\th}\Gamma^A\Gamma_{BC}\th\omega_M^{BC}\partial_i
 X^M \cdot \bar{\th}\Gamma_A\sigma\Gamma_{DE}\th\omega_N^{DE}\partial_jX^N
\Biggr]\,, \nn \\
\label{WZ:M}
S_{\rm WZ}^{\mathcal{M}} &\equiv& -2i\int\!\!d^2\sigma\,\epsilon^{ij}
\Biggl[
- \frac{1}{24}\bar{\th}\Gamma_A\sigma\mathcal{M}^2D_i\th\partial_j X^M e_M^A
\Biggr]\,.
\end{eqnarray}
This is the covariant Wess-Zumino term, which will be used for a
covariant classification of D-branes in the AdS$_5\times S^5$ background
in the next section.  The sketch of our strategy is as follows: We will
consider surface terms of $\kappa$-variation of the above Wess-Zumino
term, and investigate the boundary conditions under which all of the
surface terms vanish. Then these boundary conditions lead to a
classification of possible D-branes. 
It is worth mentioning that the surface 
terms 
do not cancel out each other.
Hence, without loss of generality, we can examine the $\kappa$-variation
of surface terms separately in each part.

\section{Classification of Supersymmetric D-branes of AdS String}

Here let us investigate the boundary conditions under which 
the $\kappa$-invariance holds. Then we will classify D-brane
configurations preserving half of supersymmetries. 

\subsection{Boundary Conditions of Covariant String}

Before going to the concrete analysis, we need to introduce the 
boundary conditions. The open-string world-sheet $\Sigma$ 
has one-dimensional boundary $\partial \Sigma$. We can impose the
Neumann condition and Dirichlet condition on this boundary $\partial\Sigma$. 
These boundary conditions are represented by 
\begin{eqnarray}
&& \partial_{\bf n} X^{\overline{A}} \equiv 
\partial_{\bf n}X^{M}e_{M}^{\overline{A}} = 0 \qquad 
\mbox{(Neumann condition)}\,, \\
&& \partial_{\bf t} X^{\underline{A}} \equiv 
\partial_{\bf t}X^{M}e_{M}^{\underline{A}} = 0 \qquad 
\mbox{(Dirichlet condition)}\,, 
\end{eqnarray}
where we have used the overline as $\overline{A}_i~(i=0,\ldots,p)$ for
the indices of Neumann coordinates and 
the underline as $\underline{A}_j~(j=p+1,\ldots,9)$
for the indices of Dirichlet coordinates. 
The operator $\partial_{\bf t}$ is a
tangential derivative $\partial/\partial\tau$ and $\partial_{\bf n}$ 
is a normal derivative $\partial/\partial\sigma$ on
the boundary $\partial\Sigma$, 
respectively. We impose the boundary condition on
the fermionic variable $\th$:
\begin{eqnarray}
P^{\pm}\th = \th\,, \quad P^{\pm} = \frac{1}{2}\left(\mathbb{I} + M\right)\,.
\label{P}
\end{eqnarray}
The gluing matrix $M$ is a product of the $SO(1,9)$ gamma matrices tensored
with Pauli matrices, and 
its concrete form will be fixed
so that the $\kappa$-variation surface terms vanish.

\subsection{D-branes of AdS Strings at the Origin}

In this subsection,
we examine the $\kappa$-variation of $S^0_{\rm WZ}$.
The $\kappa$-variation is defined by $\delta_\kappa E^A=0$, i.e.
\begin{eqnarray}
\del_{\kappa}X^{M} = -i\bar{\th}\Gamma^{M}\del_{\kappa}\th + O(\th^4)\,.
\end{eqnarray}
One finds that the $\kappa$-variation of
$S_{\rm WZ}^0$ is 
\begin{eqnarray}
&& \del_{\kappa}S^0_{\rm WZ} = \int_{\partial\Sigma}\!\!\!d\xi\,
\Biggl[
- i\bar{\th}\Gamma_{\overline{A}}\sigma\del_{\kappa}\th\cdot
\partial_{\bf t}X^Me_M^{\overline{A}} 
+ \frac{1}{2}\left(
\bar{\th}\Gamma^{\underline{A}}\del_{\kappa}\th\cdot\bar{\th}
\Gamma_{\underline{A}}\sigma +
\bar{\th}\Gamma_{\underline{A}}\sigma\del_{\kappa}\th\cdot\bar{\th}
\Gamma^{\underline{A}}\right)\partial_{\bf t}\th \nn \\
&& \qquad\qquad  - \frac{\lambda}{4}
\Bigl\{2\Bigl(
\bar{\th}\Gamma_{\overline{A}}\widehat{\Gamma}_{\overline{B}}
\sigma i\sigma_2\th\cdot\bar{\th}\Gamma^{\overline{B}}\del_{\kappa}\th 
- \bar{\th}\Gamma_{\overline{B}}
\widehat{\Gamma}_{\overline{A}}\sigma i\sigma_2\th\cdot\bar{\th}
\Gamma^{\overline{B}}\del_{\kappa}\th\Bigr)  \\
&& \qquad\qquad  +
\Bigl(\bar{\th}\Gamma^{\overline{B}}\del_{\kappa}\th\cdot
\bar{\th}\Gamma_{\overline{B}}\widehat{\Gamma}_{\overline{A}}\sigma
i\sigma_2\th - \bar{\th}\Gamma^{\underline{B}}
\widehat{\Gamma}_{\overline{A}}i\sigma_2\th\cdot\bar{\th}\Gamma_{\underline{B}}
\sigma\del_{\kappa}\th\Bigr)
\Bigr\}\partial_{\bf t}X^Me_M^{\overline{A}}
\Biggr] + O(\th^6)\,.\nn
\end{eqnarray}
First, let us consider the first line. In order for this line to vanish, 
the following two conditions have to be satisfied:
\begin{eqnarray}
\bar{\th}\Gamma_{\overline{A}}\sigma\del_{\kappa}\th = 0\, \quad 
\mbox{and} \quad \bar{\th}\Gamma^{\underline{A}}\del_{\kappa}\th = 0\,.
\label{condition:0}
\end{eqnarray}
We find that the condition (\ref{condition:0})
is satisfied if
we define the projection operator (\ref{P}) with
gluing matrices
\begin{eqnarray}
\label{cond1}
&& M = \left\{
\begin{array}{l}
m\otimes i\sigma_2\,, \quad d=2~(\mbox{mod}~4)\quad p=-1,3,7 
\\
m\otimes \rho\,, \quad d=4~(\mbox{mod}~4)\quad p=1,5,9
\end{array}
\right.\,, \\
&& m = s\Gamma^{\underline{A}_1}\cdots\Gamma^{\underline{A}_d}\,, \quad 
 s = \left\{
\begin{array}{l}
1 \quad \mbox{for $X^0$: Neumann} \\ 
i \quad \mbox{\,for $X^0$: Dirichlet}
\end{array}
\right.\,,~~
\rho=\left\{
  \begin{array}{ll}
   \sigma_1    & \mbox{when}~~ \sigma=\sigma_3    \\
   \sigma_3    & \mbox{when}~~ \sigma=\sigma_1   \\
  \end{array}
\right.\,,
\nonumber
\end{eqnarray}
where the inclusion of the factor $i$ for the case that the time
direction $X^0$ is a Dirichlet one is equivalent to performing the Wick
rotation and considering in the Euclidean formulation.  This procedure
does not affect the analysis below.  These conditions imply that, for
the $\sigma=\sigma_3$ case, D$p$-branes exist for the values
$p=-1,1,3,5,7,9$, while if we consider the $\sigma=\sigma_1$ case, then
D$p$-branes with $p=1$ and $5$ are replaced with F1- and NS5-branes
respectively\footnote{ For $p=9$, the boundary condition is
$\theta_1=\theta_2$ when $\sigma=\sigma_3$, while $\theta_2=0$ when
$\sigma=\sigma_1$.  }.  Hence, we obtain the well-known conditions for
branes in type IIB string theory in flat spacetime.  It is not clear
whether the S-duality holds for superstrings in the AdS$_5\times S^5$
and pp-wave or not, but it should exist for branes even in these
backgrounds.  Therefore, we will discuss the D-string case in parallel
as well as the fundamental string in the following consideration.

Next, we consider the second and third lines. Because these lines are
proportional to the parameter $\lambda$ characterizing the AdS
geometry, these give conditions intrinsic to the branes in the
AdS$_5\times S^5$ background.  In order to make these lines vanish, we
have to impose two additional conditions:
\begin{eqnarray}
&& \bar{\th}\Gamma_{\overline{A}}\widehat{\Gamma}_{\overline{B}}\sigma
 i\sigma_2
\th = \bar{\th}\Gamma^{\underline{B}}
\widehat{\Gamma}_{\overline{A}}i\sigma_2\th =0\,.
\label{pp2}
\end{eqnarray} 
Let us classify the configurations satisfying the
boundary conditions (\ref{pp2})\,. 

For the $d=2$ (mod 4) case,
the conditions (\ref{pp2}) are satisfied for the following case: 
\begin{itemize}
 \item The number of Dirichlet directions in the AdS$_5$ coordinates
       $(X^0,\cdots,X^4)$ is even, and the same condition is also satisfied for
       the $S^5$ coordinates $(X^5,\cdots,X^9)$.
\end{itemize}
On the other hand, for the $d=4~({\rm mod}~4)$ case, 
the following conditions should be imposed:  
\begin{itemize}
 \item The number of Dirichlet directions in the AdS$_5$ coordinates
       $(X^0,\cdots,X^4)$ is odd, and the same condition is also satisfied for
       the $S^5$ coordinates $(X^5,\cdots,X^9)$.
\end{itemize}
These conditions restrict 
the
directions to which a brane world-volume can extend,
while (\ref{condition:0}) restricts the dimension of the world-volume.
The D-brane configurations satisfying
the conditions (\ref{condition:0}) and (\ref{pp2})
are summarized in Tab.\,\ref{tab1}.
\begin{table}[htbp]
 \begin{center}
  \begin{tabular}{|c|c|c|c|c|c|}
\hline
D-instanton & D-string  & D3-brane  & D5-brane & D7-brane & D9-brane \\
\hline\hline
(0,0) & (0,2),~(2,0) & (1,3),~(3,1) & (2,4),~(4,2)
& (3,5),~(5,3) & absent \\
\hline
  \end{tabular}
 \end{center}
\caption{The possible 1/2 supersymmetric D-branes in AdS$_5\times S^5$
sitting at the origin.}
\label{tab1}
\end{table}

We found that D-strings are allowed to exist as 1/2
supersymmetric objects,
which correspond to 1/2 supersymmetric D-strings
in the pp-wave background as will be seen in section 5.
On the other hand, it is known that
there exist 1/4 supersymmetric D-strings preserving 8
{\it dynamical} supersymmetries in the pp-wave \cite{BPZ}.
In subsection 3.4, we will present a 1/4 supersymmetric D-string
which is the AdS origin of the 1/4 supersymmetric D-string
found there.
We found that D9-branes are not allowed to exist.
As will be seen in section 4,
D9-branes are not also allowed in the pp-wave \cite{DP,BP}.  
Since our
discussion contains the cases that the time direction satisfies the
Dirichlet condition,
we were able to find the D-instanton configuration.

Notably, our result agrees with the classification of branes embedded
in the $AdS_5\times S^5$ background (called AdS branes) \cite{SMT}, and
thus all of D-branes obtained here possibly correspond to the AdS
branes.  Such a correspondence emerges in the case of Dirichlet branes
of an open supermembrane in the AdS$_{4/7}\times S^{7/4}$ backgrounds,
where the classification of Dirichlet branes \cite{SaYo:ads} agrees with
that of AdS branes obtained in \cite{Kim-Yee}.

We have discussed the $\kappa$-variation of 
$S_{\rm WZ}^0$ until now. We have to take account of 
the $\kappa$-variations of
$S_{\rm WZ}^{\rm spin}$ and $S_{\rm WZ}^{\mathcal{M}}$. As we will see
later, these parts have no effect on the above classification at the origin. 
Therefore we can say that the classification of the possible D-brane
configurations at the origin has been completely accomplished. 
For D-branes sitting outside the origin, 
the $\kappa$-variation of $S_{\rm WZ}^{\rm spin}$ leads
to additional conditions. 
On the other hand, $S_{\rm WZ}^{\mathcal{M}}$ does not affect the 
results in both at and outside the origin. The discussion concerning
these points will be given in the following subsection.

\subsection{D-branes of AdS Strings outside the Origin}

Here we will consider the contribution of $S_{\rm WZ}^{\rm spin}$ part. 
The $\kappa$-variation of $S_{\rm WZ}^{\rm spin}$ is written as 
\begin{eqnarray}
\del_{\kappa}S_{\rm WZ}^{\rm spin} &=& 
\frac{i}{2}\int_{\partial\Sigma} \Biggl[
- \frac{i}{4}\Bigl\{\bar{\th}\Gamma^{\overline{A}}\del_{\kappa}\th\cdot
\bar{\th}\Gamma_{\overline{A}}\sigma\Gamma_{BC}\th 
- \bar{\th}\Gamma^{\underline{A}}\Gamma_{BC}\th\cdot\bar{\th}
\Gamma_{\underline{A}}\sigma\del_{\kappa}\th
\Bigr\}\omega^{BC}_{\overline{D}}dX^{\overline{D}} \nn \\
&& \qquad - \frac{i}{2}\bar{\th}\Gamma_{\overline{A}}\sigma\Gamma_{BC}\th\cdot 
\bar{\th}\Gamma^{\overline{D}}\del_{\kappa}\th\cdot\omega_{\overline{D}}^{BC}
dX^{\overline{A}}
\Biggr]\,.
\end{eqnarray}
In order to make the above surface terms vanish, we need the following
conditions:
\begin{eqnarray}
\label{spin}
\bar{\th}\Gamma_{\overline{A}}\Gamma_{BC}\sigma\th\cdot
\omega^{BC}_{\overline{D}} =
\bar{\th}\Gamma^{\underline{A}}\Gamma_{BC}\th\cdot 
\omega^{BC}_{\overline{D}} = 0\,.
\end{eqnarray}
These conditions are trivially satisfied at the origin
(i.e., $X^{\underline{A}}=0$), while no configurations, except for
D-instantons, satisfy these conditions outside the origin.
 
We found that 1/2 supersymmetric D$p$-branes ($p>0$) 
are not allowed except for branes sitting at the origin. 
This is because the homogeneity is not manifest in the coordinate
system we took here.
The same situation emerges
in the case of branes in the pp-wave.
There, a flat brane sitting at (outside)
the origin of the Brinkmann coordinate system
is mapped to a planer (less supersymmetric curved)
 brane in the Rosen coordinate system\cite{Bain}.
So, it is expected that a 1/2 supersymmetric brane sitting outside the
origin has a non-trivial shape in general.
On the other hand, we found that 
D-instantons are allowed to sit even outside the origin as in flat
spacetime.
This may be related to the fact that
a D-instanton world-volume is a point in 
any coordinate system.
The same situation
arises in the pp-wave case as we will see later. 

We can expect less supersymmetric
D-branes such as 1/4 supersymmetric D-branes to exist even outside the
origin as discussed in Refs.\,\cite{Bain,SMT}.
We will present this type of D-string  later.

The remaining task is to examine whether the $S_{\rm WZ}^{\mathcal{M}}$
part affects the classification at and outside the origin or
not. The
$\kappa$-variation of $S_{\rm WZ}^{\mathcal{M}}$ is 
given by 
\begin{eqnarray}
\del_{\kappa}S_{\rm WZ}^{\mathcal{M}} &=& \frac{\lambda}{12}
\int_{\partial\Sigma}\Biggl[
\bar{\th}\Gamma_{\overline{A}}\sigma\widehat{\Gamma}_{\overline{B}}i\sigma_2\th
\cdot \bar{\th}\Gamma^{\overline{B}}\del_{\kappa}\th 
-
\frac{1}{2}\bar{\th}\Gamma_{\overline{A}}\sigma\Gamma_{{B}{C}}
\th\cdot\bar{\th}\widehat{\Gamma}^{{B}{C}}
i\sigma_2\del_{\kappa}\th 
\Biggr]dX^{\overline{A}}\,.
\end{eqnarray}
The above surface terms vanish under the following conditions:
\begin{eqnarray}
\bar{\th}\Gamma_{\overline{A}}\widehat{\Gamma}_{\overline{B}}\sigma
 i\sigma_2\th =
 \bar{\th}\widehat{\Gamma}^{\overline{B}\underline{C}}i\sigma_2
\del_{\kappa}\th = 0\,,
\end{eqnarray}
which
are nothing but those obtained in the analysis
of the $\kappa$-variation of the $S_{\rm WZ}^0$ part.  Thus 
no new conditions arise, and hence the $S_{\rm WZ}^0$ part
has no effect on
our classification.

As a final remark in this section, we comment on higher order terms.
Until now we have studied the D-brane configurations up to and including
fourth order in $\th$\,.  We expect that the higher order terms do not
lead to any new additional conditions.  In other words, the conditions
we have obtained in this paper are expected to be sufficient and all
surface terms will vanish under these conditions even in the full
theory including all orders of $\th$\,.  In fact, some arguments for the
higher order terms were given in Ref.\,\cite{BPZ}.

\subsection{1/4 Supersymmetric D-string} 

We can obtain 1/4 supersymmetric D-branes
in addition to
those preserving
half of supersymmetries. 
Here we present a configuration of 1/4 supersymmetric D-string
as an example.  

Since we would like to study a D1-brane configuration, 
let us consider the following gluing matrix:
\begin{eqnarray}
M = m\otimes\sigma_1\,, \quad m \equiv \Gamma^{1\cdots 8}\,.
\end{eqnarray}
in order to pass the conditions in flat space (\ref{condition:0}). 
This gluing matrix does not automatically satisfy the conditions 
(\ref{pp2}) 
specific to the AdS geometry, and so we need to impose additional 
conditions: 
\begin{eqnarray}
\widehat{\Gamma}_{0}\sigma_2\th = \th\,, \quad 
 \widehat{\Gamma}_{9}\sigma_2\th = \th\,. 
\end{eqnarray}
Here it should be noted that the above two conditions are not
independent because of the condition $\th = M\th$\,. 
We can easily rewrite these conditions as 
\begin{eqnarray}
\Gamma^{1\cdots 8}\th_1 = \th_1\,, \quad 
\Gamma^{1\cdots 8}\th_2 = \th_2\,.
\end{eqnarray}
By the use of the relations $\Gamma^{+-1\cdots 8} =
\Gamma^{11}$ and $\Gamma^{11}
\th_{1,2}=\th_{1,2}$\,, we obtain 
\begin{eqnarray}
\label{th=0}
\Gamma^{-}\Gamma^+\th_1 = 0\,, \quad \Gamma^{-}\Gamma^+\th_2 = 0\,.  
\end{eqnarray}
When the spinors $\th_1$ and $\th_2$ are decomposed as 
$\th_{1,2} = \frac{1}{2}\Gamma^{-}\Gamma^+\th_{1,2} + 
\frac{1}{2}\Gamma^{+}\Gamma^-\th_{1,2} 
\equiv \th^{(+)}_{1,2} + \th^{(-)}_{1,2}$\,, 
the conditions (\ref{th=0}) implies $\th^{(+)}_{1,2}=0$ and 
$\th_{1,2} = \th^{(-)}_{1,2}$\,. Finally, we obtain the expressions
of additional conditions: 
\begin{eqnarray}
\Gamma^+\th_{1} = \Gamma^{+}\th_2 = 0\,. 
\end{eqnarray}
This condition reduces the number of the remaining supersymmetries, 
and the D-string configuration we considered here preserves 
quarter of supersymmetries. 

Moreover, we can see that the $S^{\mathcal{M}}_{\rm WZ}$ part has no
effect on this D-string configuration since the conditions derived from
this part is identical with (\ref{pp2})\,.  We can also check that the
$\kappa$-variation of $S^{\rm spin}_{\rm WZ}$ vanishes. The two
conditions (\ref{spin}) ensure that the surface terms originating from
the $\kappa$-variation of $S^{\rm spin}_{\rm WZ}$ vanish.  By the use of the
relation $(\Gamma^{1\cdots 8}\otimes \mathbb{I}_2)\th = \th$, we can see
that the first one in (\ref{spin}) is satisfied, while the second one is
not satisfied.  However, the term
$\bar{\th}\Gamma_{\underline{A}}\sigma\del_{\kappa}\th$ that couples to
the second condition vanishes under the condition $(\Gamma^{1\cdots
8}\otimes \mathbb{I}_2)\th= \th$, and hence the surface terms from
$S^{\rm spin}_{\rm WZ}$ vanish.  Thus, we have found a 1/4
supersymmetric D-string configuration in the AdS$_5\times S^5$
background. In particular, the supersymmetries preserved by this
configuration are not changed even if it is slid from the origin.
If we consider the Penrose limit of this D-string,
then we can recover the D-string configuration originally found in the
work \cite{BPZ},
which is 1/4 supersymmetric even outside the origin.
The D-string in the AdS$_5\times S^5$ might
correspond to $R\times S^1$
 mentioned in the works 
on the brane probe analysis given in
\cite{SMT}.  It is an
interesting future work to delve into other branes preserving 1/4 or
less supersymmetries.

\section{D-branes of PP-wave Strings Revisited}

In this section we will discuss the classification of D-branes in the
pp-wave background. 
In the work \cite{BPZ}, 
the allowed 1/2 supersymmetric D-brane configurations have been
 already classified 
for the case that the light-cone directions $(+,-)$ satisfy the Neumann
conditions,
by using the same method in this
paper.
 Here we extend the discussion given in \cite{BPZ} by 
including the case that $(+,-)$-directions are 
the Dirichlet directions.  
In particular, we find the 1/2 supersymmetric D-instanton configurations at and
outside the origin.  

By taking the Penrose limit \cite{P} around a certain null geodesic in
the AdS$_5\times S^5$ background, we obtain the maximally
supersymmetric IIB pp-wave background:
\begin{eqnarray}
\label{4.1}
&& ds^2 = 2dX^+dX^- - \mu^2(X_1^2 + \cdots + X_8^2)(dX^-)^2 
+ \sum_{m=1}^8(dX^{m})^2\,, \\
&& F_{+1234} = F_{+5678} = 4\mu\,,
\end{eqnarray}
where $F_{A_1\cdots A_5}$ is a constant
Ramond-Ramond self-dual five-form flux.

Our strategy is almost the same as in the AdS case.
We define 
the projection operator (\ref{P}) with
the gluing matrix
\begin{eqnarray}
\label{cond2}
&& M = \left\{
\begin{array}{l}
m\otimes i\sigma_2\,, \quad d=2~(\mbox{mod}~4)\quad p=-1,3,7 
\\
m\otimes \rho\,, \quad d=4~(\mbox{mod}~4)\quad p=1,5,9
\end{array}
\right.\,, \\
&& m = s\Gamma^{\underline{A}_1}\cdots\Gamma^{\underline{A}_d}\,, \quad 
 s = \left\{
\begin{array}{l}
1 \quad \mbox{for $X^+$\,,~$X^-$: Neumann} \\ 
i \quad \mbox{for $X^+$\,,~ $X^-$: Dirichlet}
\end{array}
\right.\,, \quad 
\rho=\left\{
  \begin{array}{ll}
    \sigma_1   & \mbox{when}~~ \sigma=\sigma_3    \\
    \sigma_3    & \mbox{when}~~ \sigma=\sigma_1   \\
  \end{array}
\right.\,.
\nonumber
\end{eqnarray} 
We examine the $\kappa$-variation surface terms of the action (\ref{S})
in the pp-wave background.
As before, it turns out to be enough to consider
the Wess-Zumino term (\ref{WZ}).
The supervielbein and the spin connection are given in Appendix C.
We again divide the Wess-Zumino action into the three individual parts 
as follows: 
\begin{eqnarray}
S_{\rm WZ}&=&S_{\rm WZ}^\mu+S_{\rm WZ}^{\CM}+S_{\rm WZ}^{\rm spin}\,,\\
S_{\rm WZ}^\mu&=&
-2i\int_\Sigma\Bigg[~
\frac{1}{2}e^A~\bar\theta\Gamma_A\sigma
\big(
d\theta+e^-\frac{\mu}{2}(f+g)i\sigma_2\theta
+e^m\frac{\mu}{2}\widehat{\Gamma}_mi\sigma_2\theta
\big)
\nn \\&&~~~~~~\quad 
+\frac{i}{8}\bar\theta\Gamma^A
\big(
d\theta+e^-\frac{\mu}{2}(f+g)i\sigma_2\theta
+e^m\frac{\mu}{2}\widehat{\Gamma}_mi\sigma_2\theta
\big)
\nn \\&&~~~~~~~~~~~~
\times
\bar\theta\Gamma_A\sigma
\big(
d\theta+e^-\frac{\mu}{2}(f+g)i\sigma_2\theta
+e^m\frac{\mu}{2}\widehat{\Gamma}_mi\sigma_2\theta
\big)
~\Bigg]\,,
\\
S_{\rm WZ}^{\CM}&=&-2i\int_\Sigma\Bigg[~
\frac{1}{24}e^A~\bar\theta\Gamma_A\sigma\CM^2D\theta
~\Bigg]\,,
\\
S_{\rm WZ}^{\rm spin}&=&
-2i\int_\Sigma\Bigg[~
\frac{1}{2}e^A~\bar\theta\Gamma_A\sigma
 (e^m_*\frac{\mu}{2}\Gamma_m\Gamma_+\theta)
\nonumber\\&&~~~~~~
+\frac{i}{8}\bar\theta\Gamma^A
\big(
d\theta+e^-\frac{\mu}{2}(f+g)i\sigma_2\theta
+e^m\frac{\mu}{2}\widehat{\Gamma}_mi\sigma_2\theta
\big)\cdot
\bar\theta\Gamma_A\sigma (e^m_*\frac{\mu}{2}\Gamma_m\Gamma_+\theta)
\nonumber\\&&~~~~~~
+\frac{i}{8}\bar\theta\Gamma^A (e^m_*\frac{\mu}{2}\Gamma_m\Gamma_+\theta)
\cdot
\bar\theta\Gamma_A\sigma
\big(
d\theta+e^-\frac{\mu}{2}(f+g)i\sigma_2\theta
+e^m\frac{\mu}{2}\widehat{\Gamma}_mi\sigma_2\theta
\big)
\nonumber\\&&~~~~~~
+\frac{i}{8}\bar\theta\Gamma^A(e^m_*\frac{\mu}{2}\Gamma_m\Gamma_+\theta)
\cdot
\bar\theta\Gamma_A\sigma(e^m_*\frac{\mu}{2}\Gamma_m\Gamma_+\theta)
~\Bigg]\,.
\end{eqnarray}

We examine the surface terms under the $\kappa$-variation of
these three parts in turn.
The vanishing conditions of the $\kappa$-variation surface terms coming
from $S_{\rm WZ}^\mu$
lead to the classification of possible branes sitting at the origin.
The vanishing conditions of the $\kappa$-variation surface terms
of $S_{\rm WZ}^{\rm spin}$ do not affect the classification at the origin,
but lead to the additional conditions for the brane configurations
outside the origin.
The surface terms which come from $S_{\rm WZ}^\CM$
vanish for the obtained configurations
at and outside the origin. 

The $\kappa$-variation is defined by
$\delta_\kappa E^A=0$,
which means in this background
\begin{eqnarray}
\delta_\kappa X^M=-\frac{i}{2}\bar\theta\Gamma^M\delta_\kappa\theta
+\CO(\theta^4)\,.
\end{eqnarray} 
When we perform the $\kappa$-variation on $S_{\rm WZ}^\mu$, 
the surface terms are obtained as 
\begin{eqnarray}
\delta_\kappa S_{\rm WZ}^\mu&=&
-2i\int_{\partial\Sigma}\Bigg[~
\frac{1}{2}\bar\theta\Gamma_A\sigma\delta_\kappa\theta~
 dX^Me_M^A
+\frac{i}{8}
\big(
\bar\theta\Gamma^A\delta_\kappa\theta\cdot \bar\theta\Gamma_A\sigma
+\bar\theta\Gamma_A\sigma\delta_\kappa\theta\cdot \bar\theta\Gamma^A
\big)d\theta
\nonumber\\&&~~~~~~
-\frac{\mu i}{8}\bar\theta\Gamma^-\delta_\kappa\theta\cdot
 \bar\theta\Gamma_A\sigma(f+g)i\sigma_2\theta~ dX^Me_M^A
-\frac{\mu i}{8}\bar\theta\Gamma^m\delta_\kappa\theta\cdot
  \bar\theta\sigma\widehat\Gamma_mi\sigma_2\theta~
   dX^Me_M^A
\nonumber\\&&~~~~~~
-\frac{\mu i}{16}\bar\theta\Gamma^A\delta_\kappa\theta\cdot
 \bar\theta\Gamma_A\sigma
 \big(
 (f+g)i\sigma_2\theta~ dX^Me_M^-
 +\widehat\Gamma_mi\sigma_2\theta~dX^Me_M^m
 \big)
\nonumber\\&&~~~~~~
+\frac{\mu i}{16}\bar\theta\Gamma^A\sigma\delta_\kappa\theta\cdot
 \bar\theta\Gamma_A
 \big(
 (f+g)i\sigma_2\theta~ dX^Me_M^-
 +\widehat\Gamma_mi\sigma_2\theta~dX^Me_M^m
 \big)
~\Bigg]\,.
\end{eqnarray}
The vanishing conditions for the first line read 
\begin{eqnarray}
\bar\theta\Gamma^{\overline{A}}\sigma\delta_{\kappa}\theta=0
~~~\mbox{and}~~~
\bar\theta\Gamma^{\underline{A}}\delta_{\kappa}\theta=0\,,
\end{eqnarray}
which are satisfied by the projector given above.
We thus have rederived the well-known condition,
$p=$odd,
for IIB D$p$-branes in flat background.
The additional conditions
which come from the second, third and fourth lines
are
\begin{eqnarray}
\bar\theta\Gamma_{\overline{A}}(f+g)\sigma i\sigma_2\theta=
\bar\theta\Gamma_{\underline{A}}(f+g)i\sigma_2\theta=0
~~~~\mbox{or}~~~~-\in \mbox{Dirichlet}\,,
\label{condition1}
\end{eqnarray}
and
\begin{eqnarray}
\bar\theta\Gamma_{\overline{A}}\widehat{\Gamma}_{\overline{m}}
 \sigma i\sigma_2\theta=
\bar\theta\Gamma_{\underline{A}}\widehat{\Gamma}_{\overline{m}}
 i\sigma_2\theta=0\,.
 \label{condition2}
\end{eqnarray}
The condition (\ref{condition1})
is satisfied for $p=$3 (1) mod 4 cases
by one of the followings
\begin{itemize}
  \item The number of Dirichlet directions in $(X^1,X^2,X^3,X^4)$ is even (odd), and the same condition is also satisfied for $(X^5,X^6,X^7,X^8)$\,,
  \item  $-\in$ Dirichlet,
\end{itemize}
while the second condition (\ref{condition2})
is satisfied if
\begin{itemize}
  \item The number of Dirichlet directions in $(X^+,X^1,X^2,X^3,X^4)$ is even (odd), and the same condition is also satisfied for $(X^+,X^5,X^6,X^7,X^8)$\,.
\end{itemize}
From these conditions, one can classify the possible branes
in the pp-wave background.
We denote branes with the world-volume extending along
$m$ directions in (1,2,3,4) and $n$ directions in (5,6,7,8)
as $(m,n)$-branes. In addition, 
when the directions $(+,-)$
are also Neumann directions, we denote as $(+,-;m,n)$-branes. 
We summarize in Tab.\,\ref{tab2} 
the classification of 1/2 supersymmetric D-branes
sitting at the origin.  
\begin{table}[htbp]
 \begin{center}
  \begin{tabular}{|c|c|c|c|c|c|}
\hline
D-instanton & D-string  & D3-brane  & D5-brane & D7-brane & D9-brane \\
\hline\hline
  &    & $(+,-;0,2)$,  & $(+,-;1,3)$,
& $(+,-;2,4)$, &  \\
  &    & $(+,-;2,0)$, & $(+,-;3,1)$,  & $(+,-;4,2)$,  & absent \\ 
(0,0)& (0,2),~(2,0) & (1,3),~(3,1) & (2,4),~(4,2) &    & \\
\hline
  \end{tabular}
 \end{center}
\caption{The possible 1/2 supersymmetric 
configurations of D-branes sitting at the origin 
in the pp-wave background.}
\label{tab2}
\end{table}

Next we examine the surface terms of the $\kappa$-variation of
$S_{\rm WZ}^{\rm spin}$.
We show that the surface terms vanish for branes sitting at the origin,
but lead to additional conditions for branes sitting
outside the origin.
The $\kappa$-variation surface terms are
\begin{eqnarray}
\delta_\kappa S_{\rm WZ}^{\rm spin}&=&-\frac{\mu^2}{8}\int_{\partial\Sigma}
\Big[~
\bar\theta\Gamma^{\overline{A}}\delta_\kappa\theta\cdot
 \bar\theta\Gamma_{\overline{A}}\sigma\Gamma_m\Gamma_{{+}}\theta~
  dX^{{-}}
+2\bar\theta\Gamma^{-}\delta_\kappa\theta\cdot
 \bar\theta\Gamma_{\overline{A}}\sigma\Gamma_m\Gamma_{{+}}\theta~
  dX^Me_M^{\overline{A}}
\nonumber\\&&~~~~~~
-\bar\theta\Gamma_{\underline{A}}\sigma\delta_\kappa\theta\cdot
 \bar\theta\Gamma^{\underline{A}}\Gamma_m\Gamma_{{+}}\theta~
  dX^{{-}}
~\Big]~X^m,
\end{eqnarray}
which vanish when $-\in$ Dirichlet or $X^{\underline{m}}=0$.
Thus we found that for branes sitting at the origin the surface terms
vanish, while for branes sitting outside the origin
$-$-direction has to be a Dirichlet direction.
We summarize in  Tab.\,\ref{tab:outside} the possible branes sitting
outside the origin.  
\begin{table}[htbp]
 \begin{center}
  \begin{tabular}{|c|c|c|c|c|c|}
\hline
D-instanton & D-string  & D3-brane  & D5-brane & D7-brane & D9-brane \\
\hline\hline
(0,0) & (0,2),~(2,0) & (1,3),~(3,1) & (2,4),~(4,2) &  absent   &absent \\
\hline
  \end{tabular}
 \end{center}
\caption{The possible 1/2 supersymmetric 
configurations of D-branes 
sitting outside the origin in the pp-wave background.}
\label{tab:outside}
\end{table}

Finally we examine the $\kappa$-variation surface terms of
$S_{\rm WZ}^\CM$
and show that the surface terms vanish for the brane configurations
classified above.
The surface terms become
\begin{eqnarray}
\delta_\kappa S_{\rm WZ}^\CM&=&
-\frac{\mu}{24}\int_{\partial\Sigma}\Bigg[~
\bar\theta\Gamma_{\overline{A}}\sigma
 (f+g)i\sigma_2\theta\cdot\bar\theta\Gamma^-
\delta_\kappa\theta
+\bar\theta\Gamma_{\overline{A}}\sigma
 \widehat\Gamma_{\overline{m}}i\sigma_2\theta\cdot
  \bar\theta\Gamma^{\overline{m}}
\delta_\kappa\theta
\nonumber\\&&~~~~~~
+\bar\theta\Gamma_{\overline{A}}\sigma
 \Gamma_{i}\Gamma_+\theta\cdot
  \bar\theta\Gamma^{i}fi\sigma_2
\delta_\kappa\theta
-\bar\theta\Gamma_{\overline{A}}\sigma
 \Gamma_{i'}\Gamma_+\theta\cdot
  \bar\theta\Gamma^{i'}gi\sigma_2
\delta_\kappa\theta
\nonumber\\&&~~~~~~
-\bar\theta\Gamma_{\overline{A}}\sigma
 \Gamma_{mn}\theta\cdot
  \bar\theta\widehat\Gamma^{mn}i\sigma_2
\delta_\kappa\theta
~
\Bigg]~dX^Me_M^{\overline{A}}\,.
\end{eqnarray}
One can show that these surface terms vanish when
 one of the followings
is satisfied for $p=$3 (1) mod 4 cases,
\begin{itemize}
  \item $-\in$ Dirichlet, and both $(X^1,X^2,X^3,X^4)$ and $(X^5,X^6,X^7,X^8)$
contain odd (even) number of Dirichlet directions
  \item  $-\in$ Neumann, and  both $(X^1,X^2,X^3,X^4)$ and $(X^5,X^6,X^7,X^8)$
contain even (odd) number of Dirichlet directions.
\end{itemize}
These conditions are satisfied by
the brane configurations classified above.

The longitudinal D-brane configurations agree with
those found in \cite{DP, BPZ}. 
We found that there exist transverse D-branes sitting at and outside the
origin.
D-instantons preserving half of supersymmetries are allowed to freely sit 
in the pp-wave as in flat space and AdS$_5 \times S^5$.  
D-instanton correspond to a point in the pp-wave spacetime 
while the existence of 1/2 supersymmetric D-instantons sitting
outside the origin seems to be consistent to 
the homogeneity of the pp-wave geometry.

\section{Penrose Limits of D-branes of AdS String}

In this section we will consider Penrose limits \cite{P} of the possible
D-brane configurations in the AdS background obtained in section 3, and show
that all of the D-branes in the pp-wave obtained in section 4 is
recovered as Penrose limits of those in the AdS background.

The Penrose limit is taken as follows.  One makes a set of light-cone
coordinate $(X^+,X^-)$ from $(X^0,X^9)$,\footnote{If we make a pair of
light-cone coordinates only in the AdS space
or sphere, then flat spacetime is obtained as discussed 
in \cite{BFHP2}. That is, in order to obtain the pp-wave background, 
we need to construct the light-cone
coordinates by choosing one direction
from the AdS coordinates and the other direction from 
the sphere coordinates. } scales $X^+$ as $X^+\to
\Omega^2 X^+$ and then take the limit $\Omega\to 0$.  We distinguish the
cases depending on the boundary conditions of the light-cone coordinate
$(X^+,X^-)$ as $(N,N)$ for $X^\pm \in $ Neumann directions, and so
on.
For the fermionic coordinates, we scale $\theta_+\to \Omega \theta_+$
and take the limit $\Omega\to 0$.  Some relevant aspects of the Penrose
limit are explained in Appendix D.

We examine Penrose limits of (N,N)- and (D,D)-cases.   

\subsection*{Penrose Limit of D9-brane}

There are no allowed 1/2 supersymmetric D9-brane configurations 
both in the AdS$_5\times S^5$ and in the pp-wave.  
Hence, if we want to consider the correspondence of D9-branes in the AdS
and those in the pp-wave, then we need to consider the 
less
supersymmetric 
configurations of D9-branes. 

\subsection*{Penrose Limit of D7-brane}

The possible D7-brane configurations are (3,5)- and (5,3)-branes. 
Now we cannot consider (D,D)-case since there is no Dirichlet direction 
in the $S^5$ and AdS$_5$, respectively. 

First, we consider the Penrose limit of the (3,5)-branes, 
whose boundary condition
is $s\Gamma^{ab}\otimes i\sigma_2\th =\theta$\,. 
In the (N,N)-case the
resulting boundary condition after the Penrose limit is given by
\begin{equation}
s\Gamma^{ij}\otimes i\sigma_2\th = \th\,.
\end{equation}
That is, this configuration describes the (+,$-$;2,4)-type D7-branes,
which is included in the list of the allowed D-branes
in the pp-wave case.

In the same way as in the (3,5)-brane case, we can investigate the Penrose
limit of (5,3)-branes. The boundary condition after the Penrose limit
is
\begin{eqnarray}
s\Gamma^{i'j'}\otimes i\sigma_2\th = \th\,, 
\end{eqnarray}
which is the 
$(+,-;4,2)$-brane boundary condition in the pp-wave.  Thus,
two types of D7-brane configurations in the pp-wave are rederived as the
Penrose limit of D7-branes in the AdS background.

\subsection*{Penrose Limit of D5-brane}

The allowed D5-brane configurations are (2,4)- and (4,2)-branes. 

First, we consider (2,4)-type D5-branes whose boundary 
condition is $s\Gamma^{a_1 a_2 a_3 a'}\otimes \rho\th = \th$\,
with $\rho=\sigma_1$. 
When we choose the (N,N)-boundary condition,
the above condition becomes 
\begin{eqnarray}
\Gamma^{i_1i_2i_3i'}\otimes \rho\th = \th\,,
\end{eqnarray}
after the Penrose limit.
That is, we obtain $(+,-;1,3)$-type D5-brane in the pp-wave. 
If we take the (D,D)-condition, then we get the boundary condition:
\begin{eqnarray}
i\Gamma^{+-ij}\otimes\rho\th = \th\,,
\end{eqnarray}
and hence (2,4)-type D5-branes are obtained.  
The resulting $(+,-;1,3)$- and (2,4)-type D5-branes 
are possible in the pp-wave case. 

Secondly, we study (4,2)-type D5-branes. 
In this case the boundary condition is given by 
$s\Gamma^{aa'_1a_2'a_3'}\otimes\rho\th = \th$\,.
After taking the Penrose limit, we obtain, according to the choice of
light-cone coordinates, the conditions: 
\begin{eqnarray}
&& \Gamma^{ii'_1i'_2i'_3}\otimes\rho\th = \th \qquad 
\mbox{for (N,N)-case}\,, \nn \\
&& i\Gamma^{+-i'j'}\otimes\rho\th = \th \qquad 
\mbox{for (D,D)-case}\,, \nn 
\end{eqnarray}
For (N,N)- and (D,D)-cases, we obtain $(+,-;3,1)$- and (4,2)-type 
branes, respectively. These configurations are also allowed in the
pp-wave. 

In the above discussion we have assumed the fundamental string case. 
It is also possible to consider the D-string by taking $\rho=\sigma_3$\,, 
and simultaneously D5-branes are replaced with NS5-branes.  
The only difference in our consideration of the Penrose limit 
is only a choice of $\sigma$ and the 
discussion in the D-string case is exactly the  same as the above one. 

\subsection*{Penrose Limit of D3-brane}

The possible D3-brane configurations are (1,3)- and (3,1)-branes. 

The boundary condition of (1,3)-type D3-brane is given by 
$s\Gamma^{a_1\cdots a_4a'b'}\otimes i\sigma_2\th = \th$\,, 
which becomes, according to the choice of the light-cone coordinates, 
\begin{eqnarray}
&& \Gamma^{i_1\cdots i_4i'j'}\otimes i\sigma_2\th = \th \qquad 
\mbox{for  (N,N)-case}\,, \nn \\
&& i\Gamma^{+-i_1i_2 i_3 i'}\otimes i\sigma_2\th = \th \qquad 
\mbox{for (D,D)-case}\,. \nn 
\end{eqnarray}
When we consider the (3,1)-type D3-brane whose boundary condition is 
$s\Gamma^{aba'_1\cdots a_4'}\otimes i\sigma_2\th = \th$\,,
the resulting conditions after taking the Penrose limit are given by 
\begin{eqnarray}
&& \Gamma^{iji_1'\cdots i_4'}\otimes i\sigma_2\th = \th \qquad 
\mbox{for  (N,N)-case}\,, \nn \\
&& i\Gamma^{+-ii_1'i_2' i_3'}\otimes i\sigma_2\th = \th \qquad 
\mbox{for  (D,D)-case}\,. \nn 
\end{eqnarray}
Thus, we have obtained the D3-branes in the pp-wave as
Penrose limits of D3-branes in the AdS$_5\times S^5$ background.  

\subsection*{Penrose Limit of D1-brane}

We can take (0,2)- and (2,0)-branes as 
possible D1-brane configurations. In this case we cannot take 
(N,N)-type boundary condition. 

The boundary conditions for the (0,2)- and (2,0)-type D-strings are 
\begin{eqnarray}
s\Gamma^{a_1\cdots a_5a'b'c'}\otimes\rho\th = \th \quad  
\mbox{and} \quad 
s\Gamma^{a_1a_2a_3a_1'\cdots a_5'}\otimes \rho\th = \th\,, 
\end{eqnarray}
with $\rho=\sigma_1$, respectively. 
When we consider the (D,D)-type boundary condition, 
these conditions become, after taking the Penrose limit, 
\begin{eqnarray}
i\Gamma^{+-i_1\cdots i_4i'j'}\otimes\rho\th = \th \quad 
\mbox{and} \quad i\Gamma^{+-iji_1'\cdots i_4'}\otimes \rho\th =\th\,,
\end{eqnarray}
and these conditions imply (0,2)- and (2,0)-type D-string
configurations in the pp-wave, respectively. 
If we take $\rho=\sigma_3$, then (0,2)- and (2,0)-type F-string 
configurations in the pp-wave
are obtained.

\subsection*{Penrose Limit of D-instanton} 

In this case we can consider the (D,D)-type boundary condition only. 
The condition for D-instanton in the AdS$_5\times S^5$ background 
is $i\Gamma^{01\cdots 9}\otimes i\sigma_2\th = \th$\,. After taking the
Penrose limit, we obtain the following condition: 
\begin{eqnarray}
i\Gamma^{+-1\cdots8}\otimes i\sigma_2 \th = \th\,. 
\end{eqnarray}
This condition is nothing but that of D-instanton in the pp-wave. 
\medskip

\vspace*{0.5cm}
\noindent 
Finally, we summarize the results obtained above in Tab.\,\ref{fig}.

\begin{table}
 \begin{center}
{\normalsize
\boldmath
  \[
   \begin{array}{ccccc}
\hline 
 & &  \mbox{D7-branes} & &  \\
\hline
\mbox{AdS} \qquad\quad  &\multicolumn{2}{c}{(3,5)}&
\multicolumn{2}{c}{(5,3)}   \\  
\downarrow {\mbox{\scriptsize Penrose}} \qquad 
&  D^2 \swarrow & \searrow N^2 &  D^2 
\swarrow & \searrow N^2  \\
\mbox{pp-wave} \qquad\quad & - & (+,-;2,4)  & - 
& (+,-;4,2) \\
\hline
   \end{array}
  \]
  \[
   \begin{array}{ccccc}
\hline 
 & &  \mbox{D5-branes} & &  \\
\hline
\mbox{AdS} \qquad\quad  &\multicolumn{2}{c}{(2,4)} &\multicolumn{2}{c}{(4,2)}  \\  
\downarrow {\mbox{\scriptsize Penrose}} \qquad 
& D^2 \swarrow  & \searrow N^2 & D^2 
\swarrow & \searrow N^2  \\
\mbox{pp-wave} \qquad\quad & (2,4) & (+,-;1,3) & (4,2) &
 (+,-;3,1) \\
\hline
   \end{array}
  \]
  \[
   \begin{array}{ccccc}
\hline 
 & &  \mbox{D3-branes} & &  \\
\hline
\mbox{AdS} \qquad\quad  &\multicolumn{2}{c}{(1,3)}&\multicolumn{2}{c}{(3,1)}  \\  
\downarrow {\mbox{\scriptsize Penrose}} \qquad 
& D^2 \swarrow  & \searrow N^2 & D^2 
\swarrow & \searrow N^2  \\
\mbox{pp-wave} \qquad\quad & (1,3) & (+,-;0,2) & (3,1) & (+,-;2,0)  \\
\hline
   \end{array}
  \]
  \[
   \begin{array}{ccccc}
\hline 
 & &  \mbox{D1-branes} & &  \\
\hline
\mbox{AdS}~ \qquad\quad  &\multicolumn{2}{c}{(0,2)} &\multicolumn{2}{c}{(2,0)} \\  
\downarrow {\mbox{\scriptsize Penrose}}~ \qquad 
& D^2 \swarrow  & \searrow N^2 & D^2 
\swarrow & \searrow N^2  \\
\mbox{pp-wave}~ \qquad\quad & (0,2) & \quad -\qquad \qquad    
& (2,0) & \quad - \qquad \qquad   \\
\hline
   \end{array}
  \]
\mbox{$-$:{\scriptsize We cannot take this boundary condition.}}}
 \end{center}
\vspace*{-0.5cm}
\caption{Penrose limit of D-branes of AdS string.}
\label{fig}
\end{table}

We also briefly comment on (N,D)- or (D,N)-cases. 
If we consider the Penrose limit in these cases, the boundary
condition becomes $\th_-=0$ for every D-brane. This condition 
implies the two types of conditions as follows: 
\begin{eqnarray}
M = \Gamma^{+-}\otimes \mathbb{I}_2\, \quad 
{\rm or} \quad  M = \Gamma^{1\cdots 8}\otimes \mathbb{I}_2\,. \nn
\end{eqnarray}
However, the boundary conditions associated with these gluing matrices
do not eliminate the $\kappa$-variation surface terms.
In fact, these matrices are not included in (\ref{cond2}).
It may be the case that we may not take 
(N,D) and (D,N)-boundary conditions.

\subsection*{Penrose Limit of 1/4 Supersymmetric D-string}

Here we will consider the Penrose limit of D-string configuration 
preserving quarter of supersymmetries. 
This D-string is a (1,1)-type D-string.  
When we consider the (N,N)-case, we obtain the $(+,-;0,0)$-type
D-string in the pp-wave. This D-string configuration is
nothing but the D-string found in the work \cite{BPZ}.
If we take the (D,D)-boundary condition, we obtain the (1,1)-type
D-string which is expected to be a 1/4 supersymmetric D-string
in the pp-wave.

\section{Conclusion and Discussion}

We have classified possible configurations of D-branes of
a superstring in the AdS$_5\times S^5$ background. In contrast to the
pp-wave case, the D-string configuration is realized as a usual 1/2
supersymmetric configuration. Notably, our classification result agrees
with that of 
AdS brane configurations of the brane probe analysis.
  We have also extended 
the classification of allowed 1/2 supersymmetric D-branes in the pp-wave 
\cite{BPZ} by including the case that both of light-cone 
directions satisfy the Dirichlet conditions. 
In particular, we have found that 1/2 supersymmetric D-instantons 
can exist at and outside the origin in both
AdS and pp-wave backgrounds.  The fact that D-instanton can survive even
outside the origin seems to reflect the homogeneity of these
backgrounds.  In addition, we have investigated Penrose limits of the
allowed D-branes in the AdS$_5\times S^5$ background, and it has been
shown that possible 1/2 supersymmetric configurations in the pp-wave
case can be recovered as Penrose limits of those in the AdS$_5\times
S^5$\,.

We have mainly examined the standard configurations preserving half of
supersymmetries,
but it is interesting to approach 1/4
or less supersymmetric D-branes of AdS string in our covariant
formulation.  In fact, as an example, we have presented a 1/4
supersymmetric D-string configuration in the AdS$_5\times S^5$
background. This configuration has been shown to reduce to the 1/4
supersymmetric D-string in the pp-wave preserving 8 dynamical
supersymmetries.  
D-branes of this type  are
called $D_+$-branes in the works \cite{SMT}.  On the other hand, the
standard D-branes are called $D_-$-branes.  In this paper we have
clarified the AdS origin of $D_-$-branes.  It may be possible to find
the AdS origin of $D_+$-branes.  Of course, we can also seek for the AdS
origin of oblique D-branes, curved D-branes and intersections of
D-branes in the pp-wave case (For a work in this direction, \cite{SZ}). 
In addition, giant gravitons\footnote{The matrix model on the pp-wave
also has supersymmetric fuzzy sphere solutions (giant graviton) \cite{BMN}. 
Giant gravitons in the pp-wave matrix model are discussed in
\cite{DSR,gg}.}  
or baryon vertex operators may be investigated 
in our covariant formulation if we treat these objects 
in the AdS$_5\times S^5$ by the use of our
formulation.
It is also interesting to examine Dirichlet boundaries
of an open D3-brane
constructed in \cite{MT:D3}
and compare the result with that obtained here.
Another direction is to study D-branes on other AdS
backgrounds such as AdS$_5\times S^5/\mathbb{Z}_{N}$ and 
AdS$_5\times T^{1,1}$ \cite{Ooguri}. We will report
all of these problems in another place in the near future \cite{future}.

It is also an interesting problem to apply our results to defect 
conformal field theories via the AdS/CFT correspondence \cite{BMN:open}. 
Recall that a stringy nature in pp-wave string theories 
played an important role in studies of the AdS/CFT correspondence at
the stringy level beyond supergravity approximation \cite{BMN}. Our
classification results and our methods are expected to promote the 
understanding of AdS/CFT correspondence.

\vspace*{0.7cm}

\noindent 
{\bf\large Acknowledgments}
\vspace*{0.5cm}

We would like to thank Machiko Hatsuda and Katsuyuki Sugiyama 
for useful discussion.

\vspace{10mm}

\appendix 

\noindent 
{\large\bf  Appendix}
\section{Notation and Convention}

In this place we will summarize miscellaneous notation and 
convention used in this paper. 

\subsection*{Notation of Coordinates}

For the superstring in the ten-dimensional curved space-time: 
AdS$_5\times S^5$ and pp-wave backgrounds, 
we use the following notation of supercoordinates for its superspace 
$(X^M,\th^{\bar{\al}})$: 
\begin{eqnarray}
&& M = (\mu, \mu')\,,\quad 
\mu =(0,1,2,3,4) \in AdS_5\,, \quad 
\mu' =(5,6,7,8,9) \in S^5\,, \nn
\end{eqnarray}
and the background metric is expressed by $G_{MN}$. 
The coordinates in the Lorentz frame are 
denoted by $(X^A,\th^{\al})$: 
\begin{eqnarray}
&& A=
(a,a')\,, \quad 
a=0,1,2,3,4\,, 
\quad 
a'=5,6,7,8,9 \quad \mbox{for AdS case}\,,
\nn \\
&& A=
(+,-,i,i')\,, \quad 
i=,1,2,3,4\,, 
\quad 
i'=5,6,7,8,~~~
m=(i,i') \quad \mbox{for pp-wave case}\,,
\nn 
\end{eqnarray}
and its metric is described by $\eta_{AB} =
\mbox{diag}(-1,+1,\ldots,+1)$\,
with $\eta_{00}=-1$.
The light-cone coordinates are defined by 
$
 X^{\pm} \equiv \frac{1}{\sqrt{2}}(X^9 \pm X^{0}) 
$\,. 
The coordinates of world-sheet
are parameterized by $(\sigma^1,\sigma^2) = (\tau,\sigma)$\,. 
The induced metric on the world-sheet is represented by 
$g_{ij}~(i,j=1,2)$\,. 

\subsection*{SO(1,9) Gamma Matrices}

We denote two 16-component Majorana-Weyl spinors as $\th_1$ and $\th_2$, and 
the $SO(1,9)$ Clifford algebra is written as 
\begin{eqnarray}
\{\Gamma^A,\Gamma^B\} = 2\eta^{AB}\,, \quad \{\Gamma^M,\Gamma^N\} = 2G^{MN}\,, 
\quad \Gamma^A\equiv e^A_M\Gamma^M\,, \quad 
\Gamma^M \equiv e^M_A\Gamma^A\,, \nn 
\end{eqnarray}
where the light-cone components of the $SO(1,9)$ gamma matrices 
are 
\begin{equation}
\Gamma_{\pm} = \frac{1}{\sqrt{2}}\left(\Gamma_9 \pm \Gamma_0\right)\,, \quad 
\{\Gamma^+,\Gamma^-\} = 2\mathbb{I}_{16}\,.
\nn
\end{equation}
The chirality operator and the Dirac conjugate are 
defined as 
\begin{eqnarray}
\Gamma^{11} \equiv \Gamma_{0\cdots 9}\,, \quad \bar{\th} 
\equiv \th^{T}\mathcal{C}\,, \nn
\end{eqnarray}
where the charge conjugation matrix $\mathcal{C}$ satisfies the
relations:
\[
 \mathcal{C}^{T} = - \mathcal{C}\,, \quad 
\mathcal{C}^{-1}(\Gamma^M)^{T}\mathcal{C} = -\Gamma^{M}\,.
\]
In this paper we mainly use the 32-component representation 
$\th^{T} =(\th_1,\th_2)$\,. 

\section{Coset Construction of Supervielbein in the AdS$_5\times$S$^5$}

Here we will briefly review the coset construction of supervielbein 
in the AdS$_5\times S^5$ background.
The super-AdS$_5\times S^5$-algebra is regarded as ${\it su}(2,2|4)$,
and the AdS$_5\times S^5$ manifold
can be constructed as a coset $SU(2,2|4)/\left(SO(1,4)\times SO(5)\right)$.

The bosonic part of the super AdS$_5\times S^5$-algebra is
$so(2,4)\times so(6)$ 
given by 
\begin{eqnarray}
& [P_a,P_b] = \lambda^2 M_{ab}\,, \quad [P_{a'},P_{b'}] = 
- \lambda^2 M_{a'b'}\,, \nn &\\ 
& [P_a,M_{bc}] = \eta_{ab}P_c-\eta_{ac}P_b\,, \quad 
[P_{a'},M_{b'c'}] = \eta_{a'b'}P_{c'}-\eta_{a'c'}P_{b'}\,, \nn &\\
& [M_{ab},M_{cd}] =  \eta_{bc}M_{ad}+\mbox{3 terms}\,, \quad 
[M_{a'b'},M_{c'd'}] = \eta_{b'c'}M_{a'd'}+\mbox{3 terms}\,. \nn &
\end{eqnarray}
The parameter $\lambda$ characterizes the AdS$_5\times S^5$ 
geometry, and if we take the limit $\lambda \rightarrow 0$ then
the geometry reduces to 
the ten-dimensional Minkowski spacetime. 

Now we decompose the $SO(1,9)$ gamma matrices $\Gamma^A$'s as follows:
\begin{eqnarray}
&& \Gamma^a = \gamma^a\otimes \mathbb{I} \otimes \tau_1 \quad (a=0,1,2,3,4)
\,, \nn \\
&& \Gamma^{a'} = \mathbb{I} \otimes \gamma^{a'} \otimes \tau_2 \quad 
(a'=5,6,7,8,9)\,, \nn \\
&& \Gamma^{11} = \Gamma_0\cdots\Gamma_9 
= \mathbb{I}\otimes\mathbb{I}\otimes\tau_3\,, \nn 
\end{eqnarray}
where the $2\times 2$ matrices $\tau_i$'s are the standard Pauli matrices.
The gamma matrices of the AdS$_5$ part $\gamma^a$'s 
are
\begin{eqnarray}
\gamma^i~~(i=0,1,2,3)
~~\mbox{and}~~
\gamma^4 \equiv i\gamma_{0123}\,\nn
\end{eqnarray} 
and hence $(\gamma^4)^2 = 1$ and $i\gamma_{01234} = +1$. For the 
gamma matrices of the $S^5$ part $\gamma^{a'}$'s,
we use 
\begin{eqnarray}
\gamma^{i'}~~(i'=5,6,7,8)
~~\mbox{and} ~~
\gamma^9\equiv \gamma_{5678}\,, \nn
\end{eqnarray} 
and then $\gamma^{56789}=+1$. 
The charge conjugation matrix $\mathcal{C}$ is defined as 
\begin{eqnarray}
\mathcal{C} \equiv C \otimes C' \otimes i\tau_2\,, \nn
\end{eqnarray}
where $C$ and $C'$ are the charge conjugation matrices in the AdS$_5$
and $S^5$, respectively.  We introduce chirality projection operators
$h_{\pm} = \frac{1}{2}\left(1\pm\Gamma^{11}\right)$ and Majorana-Weyl
supercharges $Q_I~(I=1,2)$ by $Q_I h_+ = Q_I$\,.  By the use of these
quantities, the fermionic part of the super-AdS$_5\times S^5$ algebra
given in \cite{MT;algebra} is rewritten as
\begin{eqnarray}
&&\{Q_{I},Q_J\} = 2i\mathcal{C}\Gamma^A (\mathbb{I})_{IJ}h_+P_{A} 
+ i\lambda\mathcal{C}\Gamma^{ab}
\mathcal{I}
(i\sigma_2)_{IJ}h_+ M_{ab} 
- i\lambda
\mathcal{C}\Gamma^{a'b'}\mathcal{J} (i\sigma_2)_{IJ}h_+ M_{a'b'}\,, \nn \\
&& [Q_I,P_a] = + \frac{\lambda}{2}Q_{J}(i\sigma_2)_{JI}\Gamma_a \mathcal{I} \,, \quad 
[Q_I,P_{a'}] = - \frac{\lambda}{2}Q_{J}(i\sigma_2)_{JI}\Gamma_{a'} \mathcal{J} \,, 
\nn \\ 
&& [Q_I,M_{AB}] = - \frac{1}{2}Q_I\Gamma_{AB}\,, \quad 
\mathcal{I}\equiv \Gamma^{01234}\,, \quad \mathcal{J} \equiv \Gamma^{56789}\,,
\nn
\end{eqnarray}
where the Pauli matrices $(\sigma_i)_{IJ}$ act on the two-component
supercharges $Q_I$\,.

Now we consider $G = g_x g_{\th}$ defined as 
\begin{eqnarray}
&& g_x = \e^{X^aP_a + X^{a'}P_{a'}}\,, \quad 
g_{\th} = \e^{Q\th}\,, \quad Q=(Q_1,Q_2)\,, \quad 
\th = \dbinom{\th_1}{\th_2}\,. \nn 
\end{eqnarray}
The supervielbeins $E^A$ and $E^{\al}$, and super spin connection $E^{AB}$ 
are defined by the relations
\begin{eqnarray}
 G^{-1}dG &=& E^{A}P_A + \frac{1}{2}E^{AB}M_{AB} + Q_{\al}E^{\al}\,,\nn
  \\
  g_x^{-1}dg_x &=& e^AP_A + \frac{1}{2}\omega^{AB}M_{AB}\,, \nn
\end{eqnarray}
where $e^A$ and $\omega^{AB}$ are the vielbein and the spin connection of the
AdS$_5\times S^5$. We can derive the expressions for
the (super)vielbein and the (super)spin connection
by expanding the l.h.s of the above
relations.  
As a result, we obtain the following expressions
\begin{eqnarray}
E^A &=& e^A + i\bar{\th}\Gamma^A\left(
\frac{\sinh(\mathcal{M}/2)}{\mathcal{M}/2}
\right)^2D\th\,, \nn \\
E^{\al} &=& \left(\frac{\sinh\mathcal{M}}{\mathcal{M}}D\th\right)^{\al}
\,, \nn \\
E^{AB} &=& \omega^{AB} -i\lambda\bar{\th}\widehat{\Gamma}^{AB}i\sigma_2\left(
\frac{\sinh(\mathcal{M}/2)}{\mathcal{M}/2}
\right)^2 D\th\,, \nn
\end{eqnarray}
where we have introduced the quantities: 
\begin{eqnarray}
\mathcal{M}^2 &=& i\lambda\left(
\widehat{\Gamma}_Ai\sigma_2\th\cdot\bar{\th}\Gamma^A 
- \frac{1}{2}\Gamma_{AB}\th\cdot \bar{\th}\widehat{\Gamma}^{AB}i\sigma_2
\right)\,, \nn \\
D\th &=& d\th + \frac{\lambda}{2}e^A\widehat{\Gamma}_Ai\sigma_2\th 
+ \frac{1}{4} \omega^{AB}\Gamma_{AB}\th\,, \nn \\
\widehat{\Gamma}_A &\equiv& (-\Gamma_a\mathcal{I},\Gamma_{a'}\mathcal{J})\,, \quad 
\widehat{\Gamma}_{AB} \;\equiv\;(-\Gamma_{ab}\mathcal{I},\Gamma_{a'b'}\mathcal{J})\,.
\nn 
\end{eqnarray}
The vielbein and the spin connection 
of the AdS$_5\times S^5$ background 
are given by 
\begin{eqnarray}
&& e^a = \left(\frac{\sinh Y}{Y}dX\right)^a\,,\quad 
e^{a'} = \left(\frac{\sinh Y'}{Y'}dX'\right)^{a'}\,, \nn \\
&& \omega^{ab} = -\lambda^2X^{[a}\left(\frac{\sinh(Y/2)}{Y/2}dX\right)^{2~b]}\,, \quad 
\omega^{a'b'} = + \lambda^2X^{[a'}\left(\frac{\sinh(Y'/2)}{Y'/2}dX
\right)^{2~b']}
\,, \nn \\
&& \quad Y^2{}^{a}_b = \lambda^{2}\left(X^2\delta^a_b - X^aX_b\right)\,, \quad 
Y'{}^2{}^a_b = -\lambda^2\left(X'{}^2\delta^{a'}_{b'} -
			  X^{a'}X_{b'}\right)\,. \nn
\end{eqnarray}
The above expressions of the spin connection are used in considering the 
D-branes sitting outside the origin. 

\section{Coset Construction of Supervielbein in the PP-wave
}

The super-pp-wave algebra is
generated by momenta, $P_-$ and $P_m$, boost, $P_m^*$, Lorentz generators of $SO(4)\times
SO(4)$, $M_{rs}$ and $M_{r's'}$, and supercharge, $Q_I$, as
\begin{eqnarray*}
&&[P_-,P_m]=P_m^*\,, \quad 
[P_-,P_m^*]=-\mu^2P_m\,, \quad 
[P_m^*,P_n]=-\mu^2\delta_{mn}P_+\,,
\\&&
[M_{mn},P_p]=\delta_{np}P_m-\delta_{mp}P_n\,,\quad 
[M_{mn},P_p^*]=\delta_{np}P_m^*-\delta_{mp}P_n^*\,,\\&&
[M_{mn},M_{pq}]=\delta_{np}M_{mq}+\mbox{3 terms}\,,\\&&
[P_r,Q_I]=-\frac{\mu}{2}Q_J\Gamma_r\Gamma_+f(i\sigma_2)_{JI}\,, \quad 
[P_{r'},Q_I]=\frac{\mu}{2}Q_J\Gamma_{r'}\Gamma_+g(i\sigma_2)_{JI}\,,\\&&
[P_-,Q_I]=\frac{\mu}{2}Q_J(f+g)(i\sigma_2)_{JI}\,,~~~
[P_m^*,Q_I]=\frac{\mu^2}{2}Q_I\Gamma_m\Gamma_+\,,~~~
[M_{mn},Q_I]=\frac{1}{2}Q_I\Gamma_{mn}\,,\\&&
\{Q_I,Q_J\}=i\CC\Gamma^+\1_{IJ}h_+~P_+
+i\CC\Gamma^-\1_{IJ}h_+~P_-
+i\CC\Gamma^m\1_{IJ}h_+~P_m
+\frac{i}{\mu}\CC\Gamma^rf(i\sigma_2)_{IJ}h_+~P_r^*\,,\\&&
~~~~~~
-\frac{i}{\mu}\CC\Gamma^{r'}g(i\sigma_2)_{IJ}h_+~P_{r'}^*
+i\mu\CC\Gamma^{rs}\Gamma_+f(i\sigma_2)_{IJ}h_+~M_{rs}
-i\mu\CC\Gamma^{r's'}\Gamma_+g(i\sigma_2)_{IJ}h_+~M_{r's'}\,,
\end{eqnarray*}
where $f\equiv\Gamma^{1234}$ and $g\equiv\Gamma^{5678}$\,.
It is known that this superalgebra is an Inonu-Wigner contraction
of the super-AdS$_5\times S^5$ algebra \cite{HKS:10} (For
AdS$_{4/7}\times S^{7/4}$ cases, see \cite{HKS:11}).
We parameterize the group manifold by
\begin{eqnarray}
G=g_xg_\theta\,, \quad g_x= \e^{X^+P_+} \e^{X^-P_-} \e^{X^mP_m}\,, \quad 
g_\theta= \e^{Q_I\theta_I}\,,\nn
\label{g:PP}
\end{eqnarray}
and define Maurer-Cartan one-forms by
\begin{eqnarray*}
G^{-1}dG=E^+P_++E^-P_-+E^mP_m+\frac{1}{2}E^{mn}M_{mn}
+E^m_*P_m^*+Q_{{\alpha}}E^{{\alpha}}\,.
\end{eqnarray*}
After simple algebraic calculation, one obtains the supervielbein 
described by 
\begin{eqnarray*}
E^A=e^A+\frac{i}{2}\bar\theta\Gamma^A\left(\frac{\sinh(\CM/2)}{\CM/2}\right)^2
D\theta\,,
\quad 
E^{{\alpha}}=
\left(\frac{\sinh\CM}{\CM}D\th\right)^\alpha\,,
\end{eqnarray*}
where we have introduced several quantities: 
\begin{eqnarray*}
&& \CM^2 =
i\frac{\mu}{2}(f+g)i\sigma_2\theta\cdot \bar\theta\Gamma^-
+i\frac{\mu}{2}\widehat\Gamma_mi\sigma_2\theta\cdot \bar\theta\Gamma^m
+i\frac{\mu}{2}\Gamma_{r}\Gamma_+\theta\cdot \bar\theta\Gamma^{r}fi\sigma_2
\\&& \qquad \quad 
-i\frac{\mu}{2}\Gamma_{r'}\Gamma_+\theta\cdot \bar\theta\Gamma^{r'}gi\sigma_2
-i\frac{\mu}{2}{\Gamma}_{mn}\theta\cdot \bar\theta\widehat{\Gamma}^{mn}
i\sigma_2\,,\\
&& D\theta = d\theta
+e^-~\frac{\mu}{2}(f+g)i\sigma_2\theta
+e^m~\frac{\mu}{2}\widehat{\Gamma}_mi\sigma_2\theta
+e^m_*~\frac{\mu^2}{2}\Gamma_m\Gamma_+\theta\,,
\\
&& \widehat\Gamma_m=(-\Gamma_r\Gamma_+f,~\Gamma_{r'}\Gamma_+g)\,,~~~
\widehat\Gamma_{mn}=(-\Gamma_{rs}\Gamma_+f,~\Gamma_{r's'}\Gamma_+g)\,,
\\
&& e^+ = dX^+-\frac{\mu^2}{2}(X^m)^2dX^-\,,~~~
e^-=dX^-\,,~~~
e^m=dX^m\,,~~~
e^m_*=X^mdX^-\,.
\end{eqnarray*}
In this parameterization, the pp-wave metric
becomes the standard form
\begin{eqnarray}
ds^2=2e^+e^-+(e^m)^2
~=~2dX^+dX^--\mu^2(X^m)^2(dX^-)^2+(dX^m)^2\,.\nn
\label{metric:PP}
\end{eqnarray}
The constructed supervielbein is used in section 4 
for the classification of D-branes 
in the pp-wave background.

\section{Penrose Limit}

It is known that the pp-wave background is obtained via the Penrose limit
from the AdS$_5\times S^5$ background.  The Penrose limit is taken by
rescaling coordinates as
\begin{eqnarray*}
X^+\to\Omega^2X^+,~~X^m\to\Omega X^m,~~\theta_+\to\Omega\theta_+\,,
\end{eqnarray*}
where $\theta_+$ is defined by $\theta_+=\CP_+\theta$ with
$\CP_+=\frac{1}{2}\Gamma_+\Gamma_-$, and then by bringing $\Omega$ to
zero.  After the Penrose limit, the scale $\lambda$ of AdS$_5$ or $S^5$
can be regarded as the scale $\mu$ of the pp-wave\footnote{The scale
$\lambda \sim 1/R$ can be absorbed and the scale $\mu$ can be introduced
by a field redefinition
\begin{eqnarray*}
X^+\to\frac{\lambda^2\Omega^2}{\mu}X^+\,, \quad 
X^-\to\mu X^-\,, \quad X^m\to\lambda\Omega X^m\,.
\end{eqnarray*}
This reveals the fact that the Penrose limit $\Omega\to 0$ is equivalent
to the limit $R\to \infty$ which is considered in the literature.}. 
Noting that the super-pp-wave algebra is an Inonu-Wigner contraction of
the super-AdS$_5\times S^5$ algebra \cite{HKS:10}, the Penrose limit of
the AdS$_5\times S^5$ group manifold parameterized by
$g_x= \e^{X^{a}P_{a}+X^{a'}P_{a'}}$ turns to be the pp-wave group manifold
parameterized by $g_x= \e^{X^{+}P_{+}+X^{-}P_{-}+X^{m}P_{m}}$.  In this
parameterization, the metric is calculated to be
\begin{eqnarray*}
ds^2&=&2e^+e^-+(e^m)^2\,,\\
e^+&=&dX^+ + \left\{1-\frac{\sin\mu X^-}{\mu X^-}\right\}\frac{X_m}{(X^-)^2}
 (X^-dX^m-X^mdX^-)\,,
\\
e^-&=&dX^-\,,
\\
e^m&=&\frac{\sin\mu X^-}{\mu X^-}dX^m
 +\left\{1-\frac{\sin\mu X^-}{\mu X^-}\right\}\frac{X^m}{X^-}dX^-\,.
\end{eqnarray*}
In order to make the metric to be of the standard form (\ref{4.1}),
we perform the coordinate transformation
\begin{eqnarray*}
\hat X^+=X^+
 +\frac{\mu}{2}\frac{\mu X^--\sin\mu X^- \cos\mu X^-}{\sin^2\mu X^-}
  (\hat X^m)^2\,,~~~
\hat X^-=X^-\,,~~~\hat X^m=\frac{\sin\mu X^-}{\mu X^-}X^m\,,
\end{eqnarray*}
under which the metric becomes the standard form
\begin{eqnarray*}
ds^2=2d\hat X^+d\hat X^--\mu^2(\hat X^m)^2(d\hat X^-)^2+(d\hat X^m)^2\,.
\end{eqnarray*}
This coordinate transformation reveals
the fact that $X^m=0$ is mapped to $\hat X^m=0$
and $X^+=X^-=0$ corresponds to 
$\hat X^+=\hat X^-=0$\,.

\end{document}